\documentclass[journal]{IEEEtran}
\usepackage{graphicx} 
\usepackage{cite}
\usepackage{amsmath,graphicx}
\usepackage{amssymb,amsfonts}
\usepackage{xcolor}
\usepackage{url}

\title{Beyond Answers: How LLMs Can Pursue \\Strategic Thinking in Education}

\author{Eleonora Grassucci, Gualtiero Grassucci, Aurelio Uncini,~\IEEEmembership{Senior~Member,~IEEE},\\and Danilo Comminiello,~\IEEEmembership{Senior~Member,~IEEE}\thanks{E. Grassucci, D. Comminiello, and A. Uncini are with the Department of Information Engineering, Electronics and Telecommunications of Sapienza University of Rome, Italy. G. Grassucci is with the Liceo Scientifico G.B. Grassi, Latina, Italy. Corresponding author's email: eleonora.grassucci@uniroma1.it.}
}

\begin{document}

\maketitle
%
\begin{abstract}
Artificial Intelligence (AI) holds transformative potential in education, enabling personalized learning, enhancing inclusivity, and encouraging creativity and curiosity. 
In this paper, we explore how Large Language Models (LLMs) can act as both \textit{patient tutors} and \textit{collaborative partners} to enhance education delivery. As tutors, LLMs personalize learning by offering step-by-step explanations and addressing individual needs, making education more inclusive for students with diverse backgrounds or abilities. As collaborators, they expand students’ horizons, supporting them in tackling complex, real-world problems and co-creating innovative projects.
However, to fully realize these benefits, LLMs must be leveraged not as tools for providing direct solutions but rather to guide students in developing resolving strategies and finding learning paths together. Therefore, a strong emphasis should be placed on educating students and teachers on the successful use of LLMs to ensure their effective integration into classrooms. Through practical examples and real-world case studies, this paper illustrates how LLMs can make education more inclusive and engaging while empowering students to reach their full potential. 
\end{abstract}
%

\section{Introduction}


Artificial Intelligence (AI) has become an integral part of people's daily lives, influencing a massive number of life aspects, from human interaction with technology to several jobs. Today workforce is facing a new reality: while AI may not directly replace jobs, workers proficient in AI will undoubtedly replace those who are not \cite{Schmelzer2024Forbes}. Thus, integrating AI into education is not only an opportunity but a necessity to equip students with the skills and knowledge needed to understand an AI-driven world. In this context, the dual relationship between AI and education blows up. On one hand, education plays a pivotal role in preparing future generations for a world where AI will be omnipresent. On the other hand, AI can be a formidable tool for enhancing education, leading towards a personalized and inclusive education, as well as strengthening students' soft skills or critical thinking and enhancing co-creativity. 

\begin{figure}
    \centering
    \includegraphics[width=0.9\linewidth]{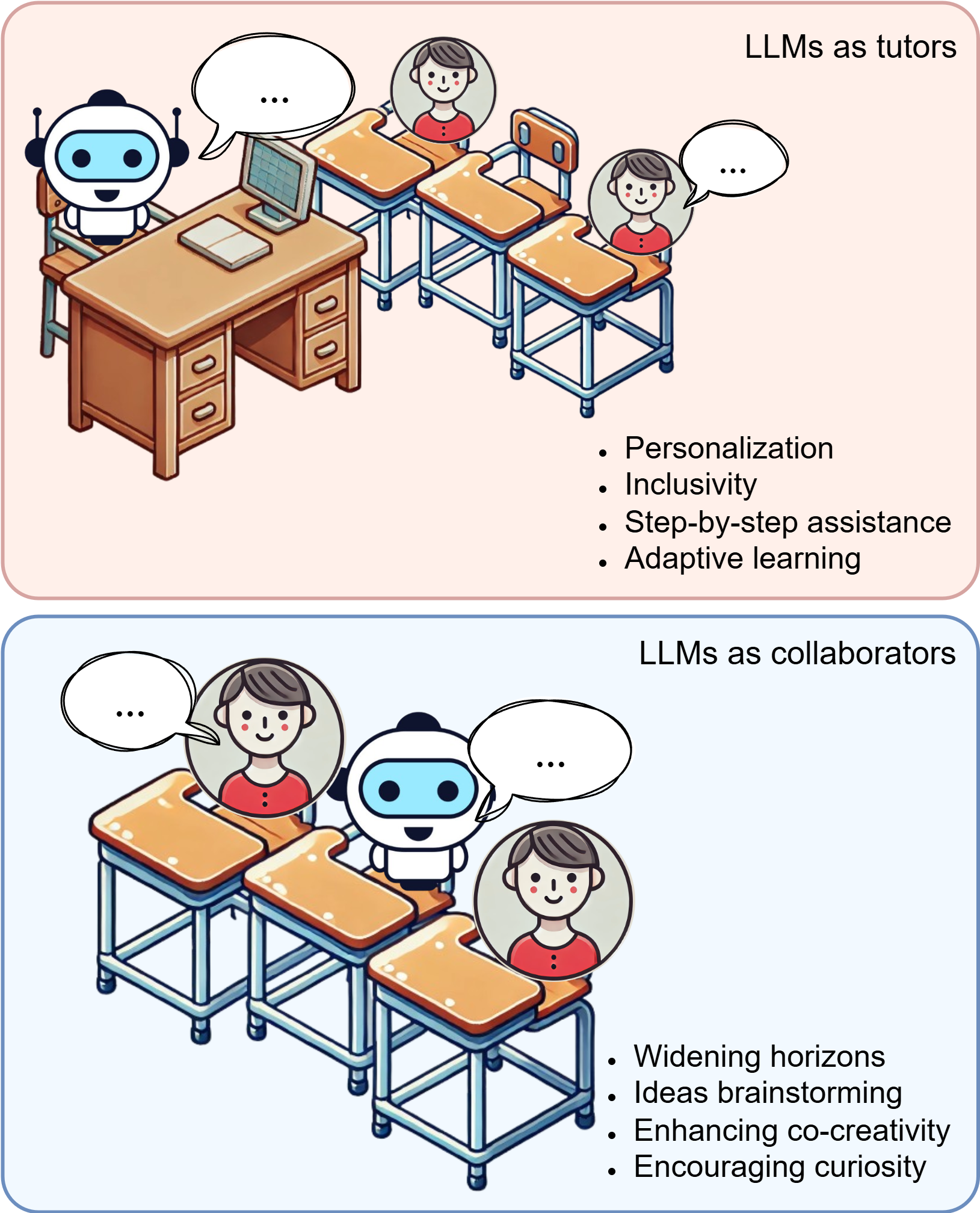}
    \caption{Representation of the dual role of LLMs in education, as tutors and as collaborators. In both the scenarios, it is crucial that students interact with the LLM not to achieve the solution but to pursue resolving strategies and a deeper understanding of concepts.}
    \label{fig:fig1}
\end{figure}

Among the various AI tools, Large Language Models (LLMs) like ChatGPT \cite{openai2023gpt4}, Claude\footnote{Anthropic. Claude 3 haiku: our fastest model yet, 2024. Available at: \url{https://www.anthropic. com/news/claude-3-haiku.}}, or Gemini \cite{geminiteam2024geminifamilyhighlycapable}, stand out due to their ability to interact with students and teachers in natural language. Unlike traditional educational tools, LLMs can provide real-time, adaptive feedback, enabling students to refine their understanding dynamically.
The integration of these tools into education may bring several key advantages that can transform how students learn and engage with knowledge and technology.
One of the most transformative benefits of integrating LLMs into education is personalization.
These models tailor responses to individual students’ strengths and weaknesses, offering customized guidance that adapts to their unique learning trajectories \cite{Wen2024kdd, Li2024BringingGA}.
However, personalization is just the first step. By adapting to individual learning needs, LLMs also create opportunities for students to explore unfamiliar and complex problems.
This helps prepare students for future works in which they will be asked to solve real-world problems and put into practice the knowledge they have learned during class.
Beyond these advantages, LLMs also encourage co-creativity by assisting students in brainstorming ideas and explore their interests through AI tools. These interactions encourage curiosity and provide students with the tools to turn their ideas into reality. However, whether guiding students through complex excercise-solving or encouraging independent inquiry, these capabilities highlight the dual role of LLMs in education, as also shown in Fig.~\ref{fig:fig1}.

\textbf{LLMs dual role in education.} While proper AI usage can provide benefits in several aspects of education, in this paper, we focus on the usage of LLMs to assist students in their learning process. We present novel perspectives on the topic and delineate possible directions that both educators and researchers can cover to exploit the potential of LLMs to enhance education delivery.
Notably, LLMs play a dual role in education: (1) as \textit{patient tutors} that guide students step by step through exercises and concepts, and (2) as \textit{collaborative partners} that help students explore problem-solving strategies and investigate novel problems. Each role serves a distinct purpose in enhancing student learning.

Acting as a tutor, an LLM provides personalized, on-demand guidance by helping students review exercises, identify errors, fill learning gaps, and deepen their understanding of concepts.
Indeed, unlike a static textbook, an LLM can respond dynamically to each student's needs, offering tailored explanations and step-by-step problem-solving assistance.
Such educational personalization may produce benefits for students with diverse learning abilities, providing specialized support for diverse learners and developing personalized learning trajectories. This implies the inclusion of all kinds of students in the class learning journey \cite{Xu2024FoundationMF}.

Beyond tutoring, LLMs can act as collaborative problem-solving partners. Instead of guiding students through predefined exercises or topics, they help students tackle new problems, explore multiple solution pathways, and critically assess different approaches.
An LLM can widen students' horizons, providing support in unfamiliar problems, helping in understanding the context, and brainstorming ideas. The student can face and solve such problems by asking the LLM to suggest strategies for solutions specifying background knowledge. Therefore, together with the LLM, the student can build resolving procedures that are based on his learning level, understanding the practical implications of the theoretical tools he has studied. In this context, both LLMs and multimedia generative models can have a strong positive impact on learners in nourishing co-creativity. Indeed, such tools can enhance the creation of projects in groups of students with average skills who may think of the resolving idea but lack the tools or the skills to put it into practice. In this context, the LLM allows students to transform their ideas or curiosity into their realization.

\textbf{Delineating the student-LLM interaction.}
However, to fully realize the potential of LLMs in education, careful consideration should be conducted on their role \cite{Xiaoming2024book}. Indeed, LLMs should not merely provide answers or act as a substitute for students' efforts. On the contrary, LLMs should support students in pursuing resolving strategies for problems, not providing the output solution, thus shifting their role from passive to active. AI and students should collaborate \cite{Latif2024PhysicsAssistantAL, Nyaaba2024GenerativeAA, Wei2024iceit}, with the technology acting as a guide that supports students in navigating challenges, encouraging them to think critically and independently. Therefore, the interaction between students and AI tools, particularly LLMs, becomes crucial and beneficial if properly leveraged to encourage collaboration between students and technology, stimulating a deeper engagement with the learning process. In this context, the benefits that AI can bring are not limited to STEM subjects but are spread across all subjects \cite{latif2024systematicassessmentopenaio1preview}.

\textbf{Contributions.} In summary, this paper aims to present the active role that AI tools, particularly LLMs, \textit{can} and \textit{should} play in transforming education across various learning levels \cite{Wang2024LargeLM}. We delineate the double role that LLMs can have in education and the advantages they can bring to students and the whole education delivery. Furthermore, we provide case studies and practical guidelines for integrating LLMs into classrooms to enhance engagement and creativity. We argue that such AI tools should not merely provide answers but actively support students in developing resolving strategies, fostering critical thinking, and promoting independence.

\textbf{Paper organization.} Section \ref{sec:tech} introduces technical information on LLMs, Section \ref{sec:person} analyzes the role of LLMs for a more personalized and inclusive education delivery, Section \ref{sec:horizons} explores the potential of LLMs in widening students horizons and facing real-world problems with the tools they have learned in class, Section \ref{sec:creativity} expounds on the enhancement of curiosity and co-creativity coming from the interactions between students and LLMs, while a discussion on the necessary education to AI is carried on in Section \ref{sec:edu} and conclusions are drawn in Section \ref{sec:con}.


%
\section{Understanding LLMs}
\label{sec:tech}

Although the research on natural language processing was born in the 1950s and largely developed in the 1990s, it knew a crucial boost in 2017, with the publication of ``Attention is all you need" \cite{Vaswani2017AttentionIA}, the paper that laid the foundations for large language models. LLMs are advanced neural networks trained on vast amounts of text data to understand and generate human-like language. These models employ a transformer architecture, initially introduced in \cite{Vaswani2017AttentionIA}, which processes text by attending to different parts of the input simultaneously, capturing complex relationships between words and concepts. The transformer architecture uses self-attention mechanisms to weigh the importance of different words in context. Such a mechanism enables the model to capture extremely long-range dependencies in text, differently from any previous method.
While the term ``large" is fuzzy, as it has been associated with both BERT models \cite{Devlin2019BERTPO} containing 110M parameters and the PaLM 2 model \cite{Anil2023PaLM2T}, which contains up to 340B parameters, modern LLMs typically contain billions of parameters trained on massive text corpora through unsupervised learning approaches, as it does not rely on labels. During training, these models learn to predict the next token (word or subword) in a sequence given the previous context. Technically, when generating text, LLMs use a process called autoregressive generation, where each new token is predicted based on all previous tokens. The model generates probability distributions over its vocabulary for each position and selects tokens through sampling strategies that may vary and are continuously updated by researchers. This probabilistic nature of text generation explains why LLMs can produce different responses to the same prompt and why their outputs are plausible but not necessarily factual. Indeed, their responses are based on patterns in the training data rather than on a deep understanding of facts. As a result, they may generate text that appears plausible but is factually incorrect, a phenomenon also known as \textit{hallucination} \cite{Huang2023ASO}. Therefore, the generation of plausible but factual content, should not be intended as an error of the LLM, since it follows the correct generation process as it is designed. Nevertheless, when leveraging these tools for factual answers, the user should consider that LLMs may generate plausible answers that may not necessarily be truthfully verified. 

%
\section{Towards Personalized and Inclusive Education}
\label{sec:person}
In a diverse classroom, some students may progress more quickly through the material, while others require more time to fully understand the concepts. Teachers often struggle to manage the wide range of abilities and knowledge levels within a single classroom, making it difficult to define precise learning paths for every student.
This is where LLMs and other AI tools can play a transformative role. By leveraging their capacity for personalization, these tools enable individualized learning experiences that adapt to the strengths, weaknesses, and interests of each student, creating a more inclusive and equitable education system.
In these cases, the LLM behaves like the patient tutor that guides the students step by step in concepts understanding, and being available at any time for any question.

\begin{figure}
    \centering
    \includegraphics[width=\linewidth]{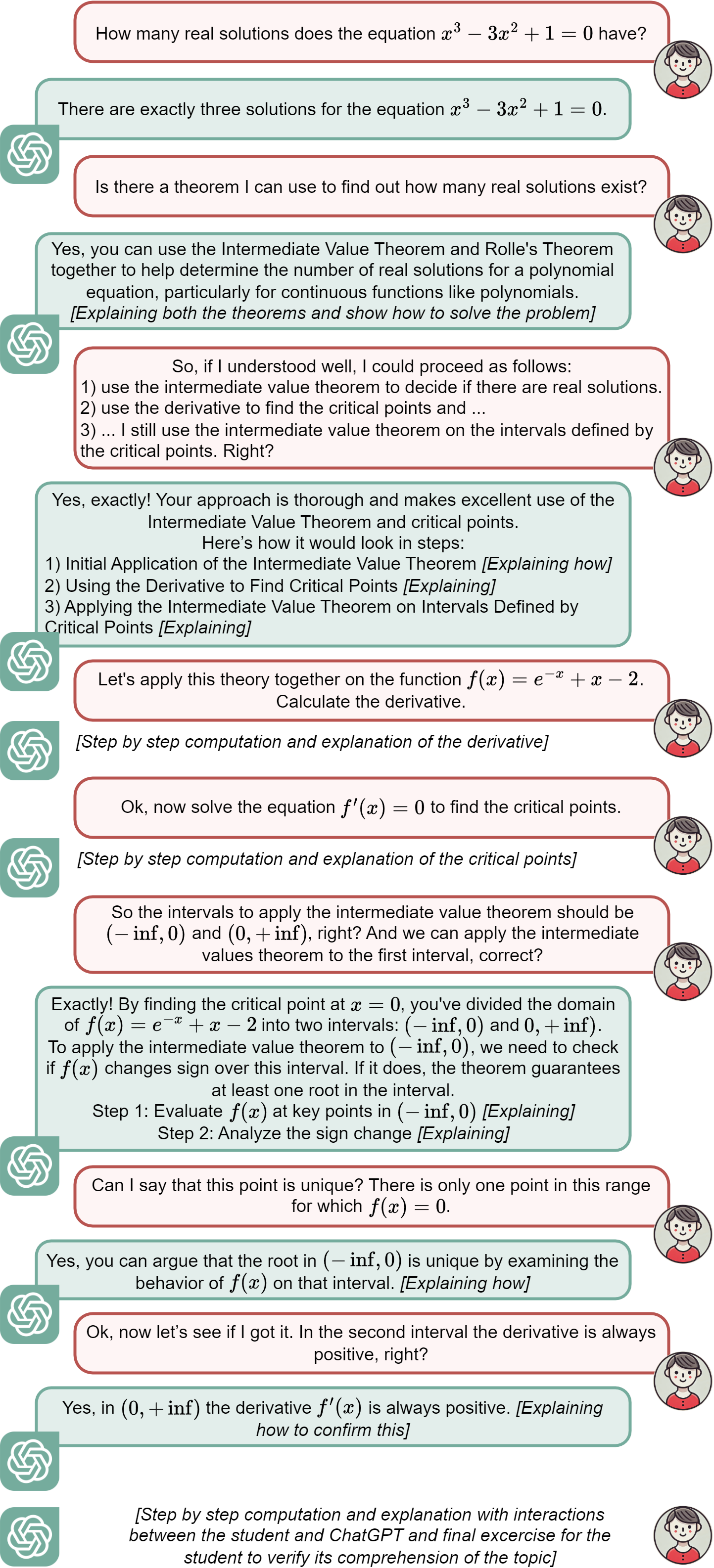}
    \caption{Example of interactive and collaborative usage of an LLM from a student trying to solve a classical math problem in a math recovery course. The student does not stop at the first answer giving the solution to his problem, but rather tries to understand the process and the theory behind the result. In the end, the student has learned the theory behind this problem being able to solve similar examples. The example is conducted with ChatGPT-4o \cite{openai2023gpt4}.}
    \label{fig:chatgpt_ex}
\end{figure}

\subsection{Tailoring learning for students with diverse abilities}

Students with diverse learning abilities often require specialized teaching approaches that address their individual challenges. LLMs can provide invaluable support by personalizing the learning experience based on the student's specific needs. For instance, students with dyslexia, ADHD, or other learning abilities might find it difficult to follow a conventional teaching approach. Through LLM platforms, these students can ask targeted questions, receive simplified explanations, or explore alternative learning paths that accommodate their challenges. Key aspects in these cases are the real-time adaptability that LLMs have, and the multimodal support they provide to diversify the learning sources and path.

\subsubsection{Real-time adaptability} LLMs can adapt their teaching strategies in real time based on student input, offering personalized explanations and alternative methods. An example is the ability of LLMs to rephrase complex instructions in simpler, more digestible language for students with cognitive impairments. For example, when tackling a signal processing problem, a student with ADHD might use an LLM to request strategic guidance on how to approach the problem in stages, with prompts encouraging focus on each step without overwhelming the student with too much information at once. Similarly, LLMs can emphasize critical parts of a lesson, such as bolding keywords or summarizing key takeaways for students who struggle with focus or attention. Furthermore, LLMs can adjust the speed of explanations and difficulty of tasks based on the student responses, ensuring a balance between challenge and comprehension. These advantages favor a more personalized education delivery, allowing students with diverse abilities to proceed at the same rate as other students.

\subsubsection{Multimodal learning support} Students with diverse abilities often benefit from multimodal approaches to learning. LLMs that accept and generate text, audio, and visual outputs can be pivotal in providing solutions to these needs. Indeed, recent LLMs have been developed to be naturally multimodal \cite{Lee2024RealizingVQ}. This means they accept diverse modality inputs ranging from text to audio and images and are able to interact with users both in natural language and in audio waveform. Such multimodality further enhances the inclusivity of students enabling them to choose the way they prefer to interact with the LLM. Students may prefer to speak rather than watch the screen or they may prefer to do everything silently just chatting with the LLM. For instance, students with dyslexia or visual impairments can leverage LLMs to read instructions, explain concepts, or narrate stories, like a real personalized tutor. Interactive dialogues enable these students to engage without relying on text-based materials. Another benefit may occur for students who struggle with abstract reasoning for which LLMs can generate visual representations of concepts, such as graphs, flowcharts, or animations, to make complex ideas more accessible.

However, the key aspect is that LLMs should not simply provide the answers to problems. Instead, they should assist students in navigating the learning process by helping them break down complex tasks into smaller, more manageable steps, as a real tutor might do. By facilitating a personalized approach, LLMs empower students with diverse learning abilities to pursue problem-solving strategies that align with their cognitive and emotional needs, encouraging them to be autonomous and gain independence and confidence.

\subsection{Classrooms of diverse-background students}

Large classrooms may be composed of students with diverse skills and backgrounds, making it difficult for teachers to design a single curriculum that effectively meets the needs of every student. In this context, students can ask LLMs questions that are specific to their understanding and receive tailored guidance on how to approach the teacher-proposed problem. Rather than relying on a uniform teaching method, LLMs allow students to explore problems from different angles, depending on their individual knowledge and learning style, thus personalizing their education path. This is particularly useful in complex subjects like maths and signal processing, where students may need to revisit foundational concepts or receive contextual information to fully grasp advanced material. Moreover, missing some basic concepts may often undermine the understanding of future topics. In such cases, acting immediately is crucial. Unfortunately, it is difficult to timely intervene with each single student's difficulties without a personal tutor. Therefore, students' issues may worsen and difficulties may further increase. The LLM could be that personal tutor for each student.

In the following subsection, we present a case study conducted with a high school student in a math recovery course in which the student tries to better understand a topic covered during class.

\subsubsection{Case Study: Maths Recovery Course}

Figure~\ref{fig:chatgpt_ex} shows an example of an exercise to be solved in a maths recovery course. The task is finding the real solutions to a third-grade equation for a student from the third year of Italian high school, grade 11 in the US. The student is tasked to solve the problem, to show the procedure and the reasoning behind the solution and also to interact with the LLM to be guided towards the solution. As shown in Fig.~\ref{fig:chatgpt_ex}, at the beginning the student directly asks the solution to the LLM, obtaining the answer without learning the procedure to obtain it. Therefore, the student asks the LLM to explain the theory behind the solution and obtains back the explained theorem. Up to this point, the interaction is still \textit{cold} between the two agents, with the student gaining generic explanations with few interactions. The real potential is unveiled in the next messages, where the student asks the LLM to guide him step by step toward the solution, letting the LLM do the complex calculus and helping the student to understand the reasoning behind the resolutive strategy. In this way, the student gains a deeper understanding of the process and of the solution, being able to solve similar exercises. In the end, the student can also ask the LLM to provide him with additional tests to verify his comprehension. 

This personalized guidance not only helps students fill knowledge gaps but also strengthens their ability to approach future problems independently.
By focusing on the resolving process rather than providing the result, the LLM encourages students to build their problem-solving skills over time, allowing them to gain confidence in their own abilities.

\subsection{Improve LLMs personalization in education}

Concurrently with the introduction of AI usage into schools and universities, novel models and methods can be developed to further enhance the personalization of education delivery.
While LLMs such as Gemini and ChatGPT possess impressive capabilities, their potential for personalization can be significantly enhanced to better cater to individual learning needs. One of the most promising directions is the incorporation of context-aware interaction. By integrating LLMs with individual learning profiles, which include student history, preferences, and goals, these tools can offer tailored responses that address specific challenges and strengths. For example, a student who frequently struggles with algebra could receive proactive, targeted explanations and practice problems, while another excelling in the same subject might be guided toward advanced applications.

Another critical area of improvement lies in enhancing the emotional intelligence of LLMs. Combining LLMs with multimodal inputs such as facial expression and tone analysis or wearable sensors that monitor physiological signals can enable these models to recognize and respond to students' emotional states \cite{Tanaka2024LeveragingLM, li2023large}. An LLM may be aware of the interlocutor's emotional status by being conditioned with signals captured by a student wearable device capturing ECG \cite{SweeneyECG, Yan2023emotion}. By making the LLM aware of the interlocutor's emotions, the model can tailor the answers according to the emotional status of the student, emulating the empathetic support of a human teacher. Examples are being aware of the attention rate of the student, if the topic is comfortable or if it causes bad feelings, and providing answers accordingly or stimulating the student's attention with some quiz or questions. Another example may be the scenario of a student preparing for a signal processing course test. If a wearable device detects elevated stress levels during a practice session, the LLM could pause the activity to offer a short, engaging game or a simpler set of questions to rebuild confidence. By resuming the original task gradually, student learning continues in a more supportive environment, highlighting how emotional intelligence in LLMs could enhance educational outcomes.

%

\begin{figure*}
    \centering
    \includegraphics[width=\textwidth]{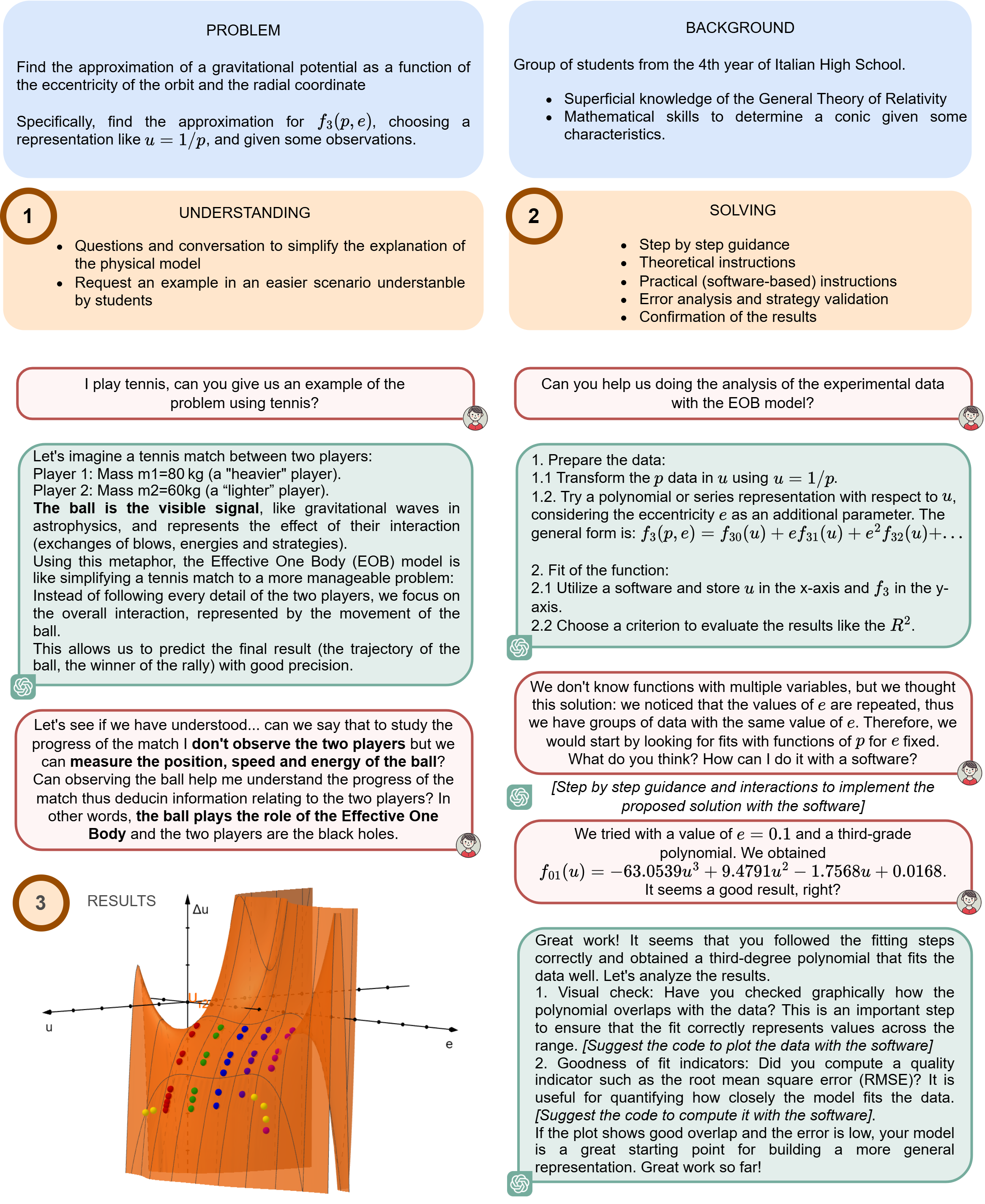}
    \caption{A group of students are asked to find the solution to a complex problem of which they do not have prior knowledge or background. First, the group interacts with the LLM to understand the problem and the suggested model to solve it (Step 1). They try to bring back the problem to an easier scenario they can better understand. Once understood, the group starts trying to find a strategy to solve the problem together with the LLM and the help of a CAS software (Step 2). In the end, the students find a strategy to solve the problem and obtain satisfactory results (Step 3). Chats are cut for visualization purposes. The example is conducted with ChatGPT-4o \cite{openai2023gpt4}.}
    \label{fig:cs_horizons}
\end{figure*}

\section{Widening Student Horizons}
\label{sec:horizons}
In many educational scenarios, students encounter problems that require a broader understanding of concepts than they have studied directly, sometimes struggling to link them and arrive at a solution. In other cases, the teacher is limited by the educational offer itself to propose real problems to students since they do not have enough skills or guidance to put into practice the theoretical concepts they have learned. In these situations, an LLM acting as a collaborator can represent a valid solution to both circumstances.

An artificial collaborator can offer significant support by helping students understand the context of unfamiliar problems and guiding them to explore potential strategies for solutions. Rather than providing an immediate answer, an LLM can prompt students to specify their background knowledge, allowing the tool to offer suggestions that are aligned with the student's level of understanding. For instance, in applied mathematics, a student may be tasked with solving a real-world problem that requires the integration of multiple theoretical concepts covered in class. If the problem feels unfamiliar, the student could leverage the LLM to request guidance on how the theoretical tools he has studied—such as calculus or linear algebra—can be applied to the specific scenario. By doing so, the student not only solidifies their grasp of the theory but also learns how to extend that understanding to solve new, more complex problems. The process of asking questions and receiving strategic guidance from LLM helps the student develop a deeper appreciation for the practical implications of their theoretical knowledge. In addition, the availability of such a collaborator allows teachers to expand their educational offer, proposing students more complex problems and asking them to solve such problems with the help of their personal collaborator. Therefore, involving LLMs as artificial colleagues indirectly helps improve and diversify the educational offer. Moreover, it prepares the student for future working scenarios, in which he will be asked to solve real-world problems with the skills he has acquired during his education process.

In the following subsection, we present the case study of a real-world research problem proposed to a group of high school students that, at the beginning, do not have the background knowledge to solve it, thus they require a step by step guidance to find a resolution strategy.

\subsubsection{Case Study: Approximate the gravitational potential}

Figure \ref{fig:cs_horizons} presents the case study of a problem proposed to high school students by Prof. D. Bini and starting from \cite{Barack2011BeyondTG}. Given a set of observations, the students are asked to find a function that fits the gravitational potential of two black holes as a function of the eccentricity of the orbit and the radial coordinate. The project is assigned to a group of students from the fourth year of Italian High School (corresponding to the grade 12 - Senior Year in the US). Given that the students do not have any prior knowledge of the problem, the role of the LLM is crucial first to translate the problem into an easier scenario. By interacting with the LLM, the students can better understand the problem and the reasoning behind the suggested resolutive model. It is interesting to note that students ask the LLM to bring the problem back into a scenario they can easily understand, which is a tennis game, as highlighted in the first chat of Fig. \ref{fig:cs_horizons}. In this way, whatever the problem is, the students can grasp the reasoning behind it in a simpler scenario. Once the students become familiar with the problem, they can start exploring possible strategies to find the required fitting function. In this phase, the students interact with the LLM for several purposes. Firstly, the LLM helps them establish a procedure to follow for their experiments. Secondly, as a collaborator, the LLM can validate their intuitions encouraging the students to pursue the strategies they propose or correcting them in the case of wrong assumptions. Additionally, the LLM can support the students in using CAS software to solve complex calculus and plot the results. Thirdly, the LLM can validate the results the students discovered, guiding them through the analysis of the results up to a robust result. It is notable that the result obtained, shown in Step 3 of Fig. \ref{fig:cs_horizons}, is a considerable result both in scientific terms and for students of high school, which usually focuses on more standard exercises to be solved by hand.

The example in Fig. \ref{fig:cs_horizons} shows the several advantages that LLMs can bring to education in widening the student's horizons. The direct consequence is that, by interacting with the LLM, the students face and solve a complex real-world problem by leveraging the skills learned during class while also learning much more complex tools and scenarios they were not aware of. However, the example in Fig. \ref{fig:cs_horizons} brings more crucial advantages. Indeed, the students obtain more self-confidence in their abilities and in the tools they learned in class, and also in the importance of basic tools in real-world scenarios. This aspect is crucial to enhance the trustworthiness in the education process, especially in recent years with the diffusion of several (uncertified or self-created) learning sources. Additionally, the LLM guides them in defining a rigorous scientific procedure, as it is clear from the right chat (Step 2) in Fig. \ref{fig:cs_horizons}. The LLM suggests to students to first \textit{Prepare the data}, and then to \textit{Fit the function}. After these steps, the LLM suggests to \textit{analyze the results} and to define an objective metric to evaluate the obtained results. These instructions guide the students through the experiment by defining a clear and reproducible procedure typical of scientific discoveries. Indirectly, this enhances scientific thinking in students and teaches them that solving a problem is not only obtaining the solution but also establishing a procedure and defining a successful and reproducible resolutive strategy. More projects like this are available at \url{http://researchinaction.it/}.




%

\section{Enhancing Co-Creativity and Curiosity}
\label{sec:creativity}
In most grades of education, from secondary school and high school to university courses, classes are composed of students with different intelligences, passions, and skills. One of the teachers' challenges is to take average and below-average students to success, as often upper-average ones will probably succeed regardless of the teacher. Additionally, whatever their abilities are, such students may be curious to discover new concepts or real-world problems and the role of the teacher is also to encourage and inspire such curiosity.
In this Section, we expound on the possibilities that AI tools, especially LLMs, can bring to help students with average skills while also inspiring upper-average students and providing tools to satisfy students' curiosity. Similar to Section \ref{sec:horizons}, here we intend the LLM as a wise collaborator that can help students put into practice their ideas or satisfy their curiosity.

\subsection{Co-creativity and curiosity in students with average skills}
In applied mathematics or signal processing real-world cases, students may have innovative ideas for solving a problem or developing a project but lack the technical expertise or tools to turn those ideas into reality. AI tools, such as generative models like text-to-image/audio diffusion models and LLMs among others, can bridge this gap by offering strategic support and suggestions that help students translate their ideas into actionable steps. For example, a group of students working on a signal processing project, that may be noise cancellation from audio signals, might conceptualize a novel approach to define nonlinear adaptive filters but lack the coding proficiency or understanding of certain advanced techniques. By interacting with the LLM, they can ask questions that clarify these techniques, receive code snippets, or even get guidance on how to refine their ideas. The LLM acts as a facilitator, helping students move beyond the initial concept stage and empowering them to implement their ideas more effectively. This process not only fosters creativity but also ensures that students remain engaged and motivated as they see their ideas come to life through collaboration and AI support. In the following case study, we present the process and the results of an activity directly proposed by students who were curious to study and understand the behavior and the spread of the COVID-19 epidemic.

\subsubsection{Case Study: Epidemics Spread} 
\begin{figure*}
    \centering
    \includegraphics[width=0.9\textwidth]{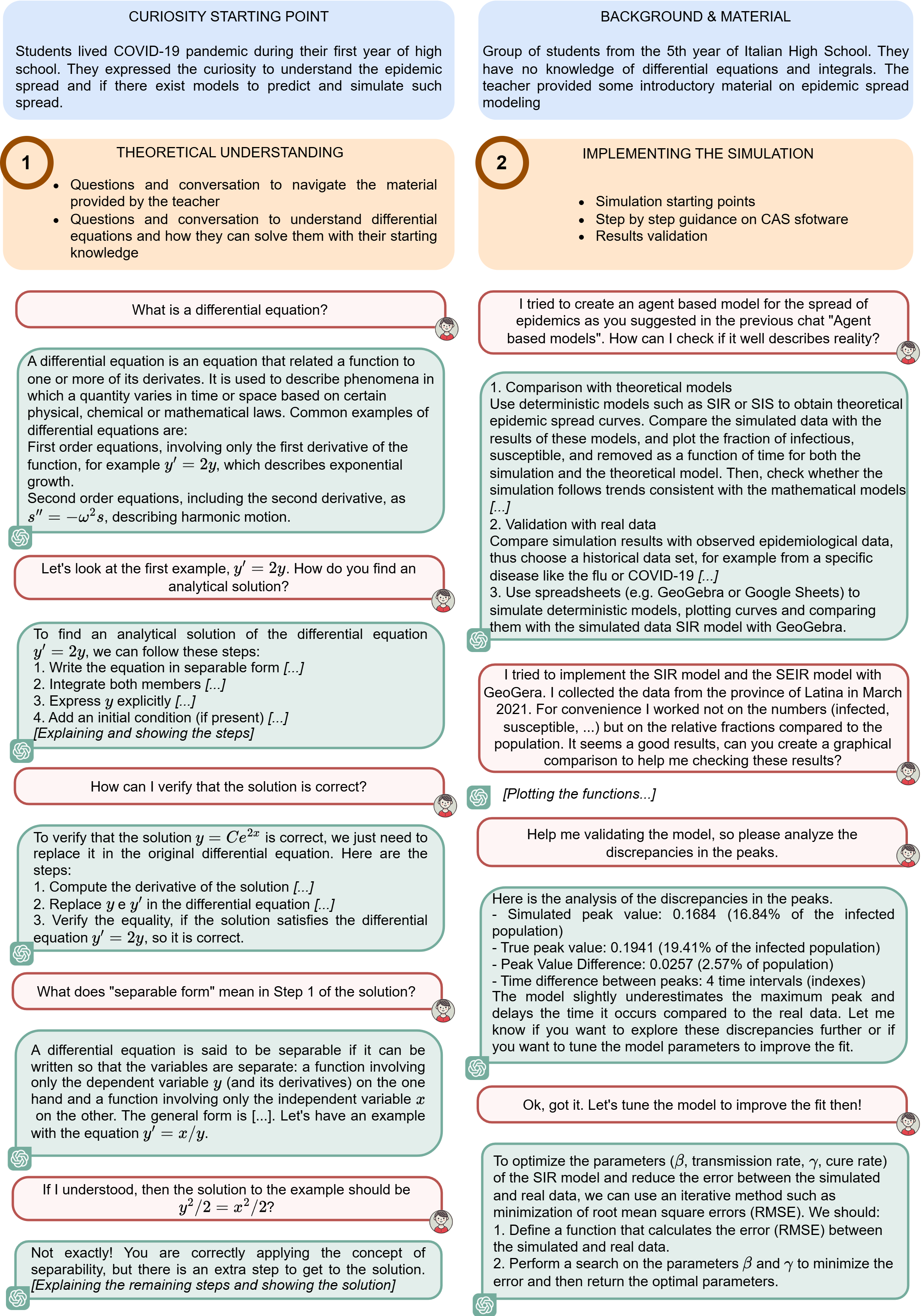}
    \caption{Guided by the curiosity to understand the epidemic spread of COVID-19, the students expressed an interest in studying epidemic spread models and realizing a simulation on the data from their Italian region. Here, at first, the LLM helps students navigate the material provided by the teacher and understand how to solve differential equations with their background knowledge. Secondly, they interacted with the LLM to realize the simulation on real-world data of their interest and estimate the COVID-19 spread. Simulation results are presented in Fig. \ref{fig:epidemic results}.}
    \label{fig:curiosity}
\end{figure*}

\begin{figure*}
    \centering
    \includegraphics[width=\textwidth]{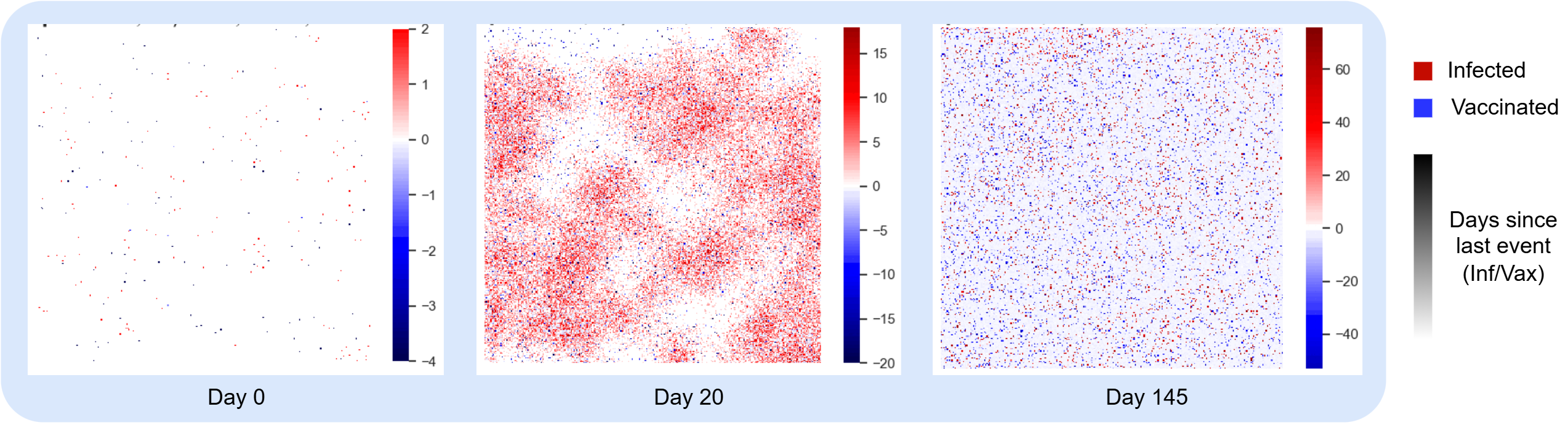}
    \caption{Results at different times (Day 0, 20, and 145) of the study developed in the case study: epidemic spread. Simulation on a population of 90000 people for 200 days, no vaccinated people at the beginning, and vax life of 50 days. Simulated 4 interactions per person per day. After just 20 days, the epidemic already significantly spread infecting a considerable number of people. After 145 days, the people started to get vaccinated and the epidemic spread crucially reduced.}
    \label{fig:epidemic results}
\end{figure*}

The activity of studying epidemic diffusion is proposed by a group of students from the last year of Italian high school. Such students lived in Italian lockdown due to the COVID-19 diffusion, which in Italy has been particularly hard and impacting on their daily lives, in their first year of high school. Figure \ref{fig:curiosity} shows the process and some examples of the chat between the students and ChatGPT, while Figure \ref{fig:epidemic results} shows the results of the simulation the students perform. Once the teacher and the expert tutors E. Montefusco and C. Rizzo provided the students the material showing the basic concepts of epics spread modeling the learning process followed two main steps. In the first step, the students asked the LLM to help them navigate the material, at first asking general questions on epidemic spread models. Successively, since such models are based on differential equations and the students did not possess knowledge about them, they started asking for information on what are differential equations and how they can solve them with the background knowledge they possess. Following the first step, in the second one, the students started implementing the simulation with real-world data. In this phase, the LLM helps them develop the simulation, suggesting to use of open-source spreadsheets (as per the teacher's instructions), and validating the results obtained by the students.
Simulation results are shown in Fig. \ref{fig:epidemic results}, where at day 0 there are very few infected, and after only 20 days, the largest part of the population has been infected. The effect of vaccines is then clear in the last plot where, even though not the whole population is vaccinated, the epidemic's spread is significantly limited. More projects like this at \url{http://researchinaction.it/}.

During this project, the students act independently, sometimes randomly, but slowly expanding their knowledge and refining their skills. The LLM, instead, behaves like an assistant. Interestingly, in this case, the LLM acts as a collaborator for both the students and the teacher. In the first case, it is an assistant helping them during the investigation of the problem, while also being a reliable assistant for the teacher, thoroughly following the original material and the instructions provided by the educator. In this case study, the interaction with the LLM guided students to the development of a simulation of an event that really happened, the COVID-19 spread, which strongly impacted students' lives. As a direct consequence, students gained a better understanding of epidemics spread and touched with their hands the impact of the epidemics and vaccines on their population. Indirectly, students gained self-confidence, starting to believe in the theoretical concepts and tools they have learned during class, in their capacity to understand complex problems, and in the applicability of such tools in real-world practical scenarios.

\subsection{Co-creativity in students with upper-average skills}
Sometimes, students with skills above the classroom average might not find enough learning incentives during class as the teacher should maintain a rate that fits the class average. Indeed, sometimes, standard curricula may leave these students under-stimulated and unable to reach their full potential. This situation can result in disengagement and a lack of motivation, preventing them from developing advanced skills or exploring innovative ideas. By introducing LLMs usage in education, this kind of students can interact with the tool to build their own projects advancing the knowledge they have gained in the class with frontal lectures. With the help of the AI collaborator, we may ensure that no student is left behind, not because of a lack of support, but due to insufficient challenges. LLMs allow upper-average students to transcend the boundaries of traditional curricula, granting them the autonomy to explore, experiment, and excel.

\begin{figure}
    \centering
    \includegraphics[width=\linewidth]{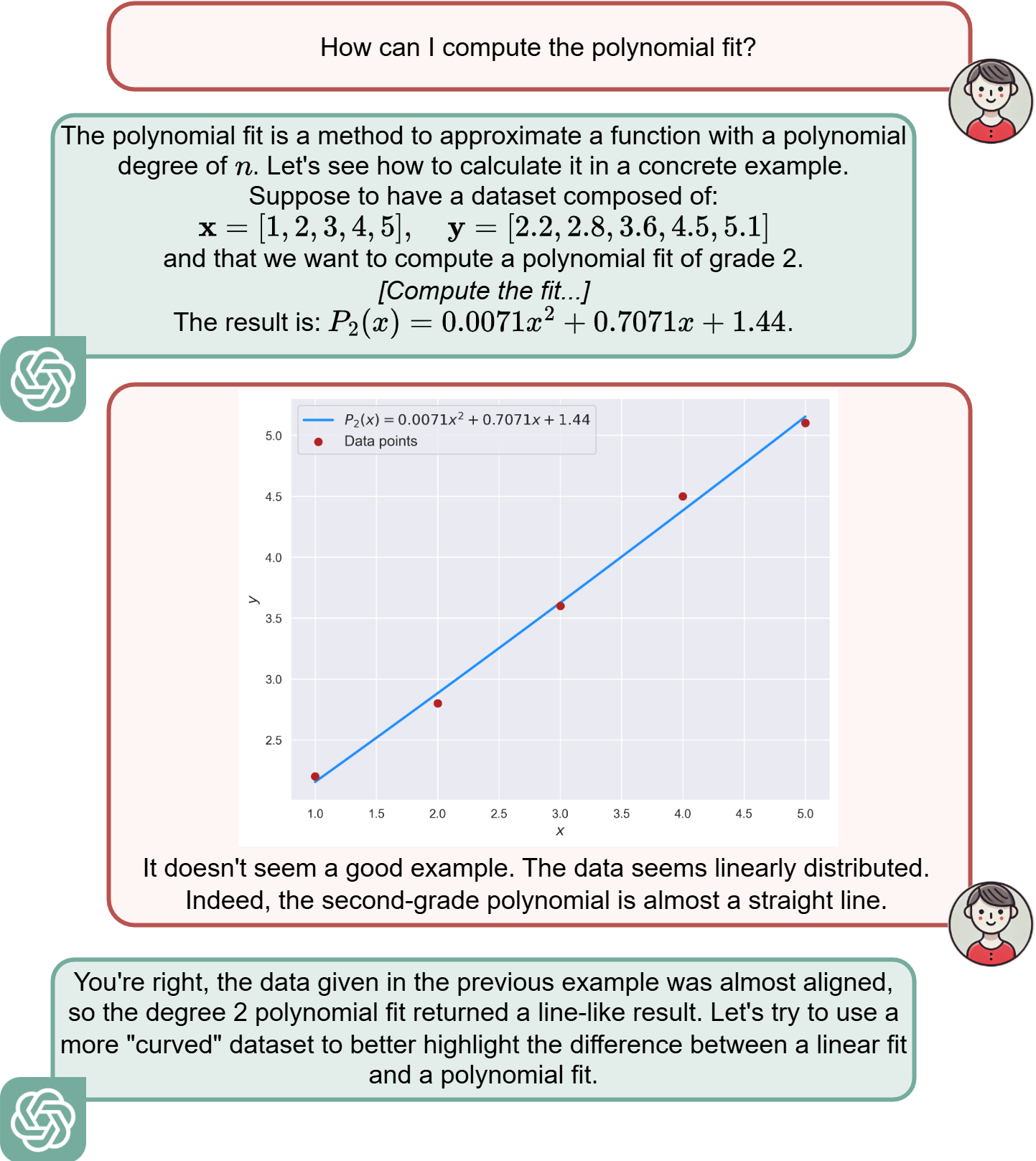}
    \caption{During the explanation, the LLM made a mistake trying to fit almost linearly aligned data with a second-grade polynomial. The group of students plotted the results and realized the error, so they came back to the LLM to ask for a better example to understand the polynomial fitting. Unintentionally, the students develop critical thinking with respect to the answers provided by the LLM, refining their soft skills. The example is conducted with ChatGPT-4o \cite{openai2023gpt4}.}
    \label{fig:error}
\end{figure}

\subsection{Refining soft skills}
To extract the most value from generative AI tools and generate better images or gain the best answer from the LLM, students must learn how to formulate effective prompts, critically evaluate the LLM-generated suggestions, and refine their strategies based on the feedback they receive. In this context, the process of working with AI is not passive, as students must actively engage with the tool, which, in turn, hones their creative and analytical abilities. Interestingly, AI is making soft skills more essential for success in both educational and professional settings. For example, students who collaborate on a group project using AI must communicate effectively with one another to formulate shared goals and design prompts that guide the AI model output toward a meaningful outcome. In this way, AI supports the development of hard skills, such as coding or problem-solving, and reinforces the necessity of soft skills in maximizing the benefits of AI-driven tools.

\subsubsection{Case Study: Guiding the LLM towards a better solution}
We recall the case study in Section \ref{sec:horizons} where the students are asked to find a good approximation of the gravitational potential given two variables. Here, we show an example of how the interaction with LLMs indirectly helps students develop critical thinking in a chat coming from the problem in Section \ref{sec:horizons}. During the process of looking for a proper fitting function given a set of observations, the students ask the LLM for an explanation about the polynomial fit, as we show in Fig. \ref{fig:error}. To provide an example, the LLM randomly generates a small set of points and fit a second-grade polynomial to those data. At this point, the group of students plots the data and the function to observe the behavior of the fitting, realizing that the points were linearly distributed, thus the nonlinear fitting function was unnecessary. This means that the LLM provides a wrong example in its explanation. However, the interesting aspect of Fig. \ref{fig:error} is that, unintentionally, the students develop critical thinking about the answers provided by the LLM, thus refining their soft skills while working wth the LLM.


\section{Education To AI}
\label{sec:edu}

In the preceding sections, we explored the substantial benefits that integrating AI tools, particularly LLMs, can bring to education. These advantages include cultivating a more inclusive learning environment for students with diverse abilities, enabling learners to address real-world problems by applying classroom knowledge, and enhancing creativity through co-creativity.
However, these transformative benefits hinge on a critical prerequisite: AI must be used as a tool to understand processes, gain insights, and develop resolution strategies, rather than as a mere shortcut to obtaining solutions. This distinction underscores the need for a well-structured education in AI usage. It is not enough to simply introduce AI tools into the learning process, students and teachers alike must be equipped with the knowledge and skills to use these tools effectively. Teachers' role should be shifted from traditional knowledge transmission to doing it together with teaching how to use AI in a thorough learning path. They should teach how to learn AI tools for diverse purposes to be used both during classes and individually. This is an urgent aspect that teachers and schools should consider, as students are already starting to use AI in their educational path. Equally important is the need to cultivate a sense of discernment and critical thinking in students regarding AI-generated outputs. LLMs, while powerful, are not infallible. Their responses are plausible but not always accurate or truthful. Thus, students must learn to critically evaluate AI outputs, cross-check information, and refine their problem-solving approaches based on the feedback received from these tools.

The integration of AI education into school curricula should not be limited to technical skills but must encompass a broader understanding of AI's societal impact. As AI continues to shape various aspects of daily life, proficiency in using and understanding AI tools is becoming increasingly necessary to navigate the complexities of the modern world.
Educational institutions should prepare students to meet the demands of future job markets, where proficiency in AI will be a decisive factor for success \cite{Schmelzer2024Forbes}. Training students in AI usage is not only about equipping them with tools for academic purposes but also about fostering adaptive and innovative thinkers who can respond to evolving challenges. AI education should aim to develop a generation that views technology as an enabler for creativity, and critical thinking, and a collaborator in finding the resolving strategy.

\subsection{Guidelines for Teachers Using LLMs in Daily Classes}
The successful integration of AI tools like LLMs into classrooms requires a strategic and thoughtful approach by educators. Teachers play a pivotal role in mediating the interaction between students and AI, ensuring that the tools are used to enhance learning rather than hinder it. Below, we outline key strategies that teachers can adopt to effectively incorporate LLMs into their daily teaching practices.

\textbf{Encourage active collaboration.} Teachers should encourage students to engage with LLMs actively rather than passively consuming information. This involves teaching students to use LLMs to explore problems, analyze concepts, and refine strategies, as we explored in previous sections. This includes that teachers should also start proposing students problems that are specifically tailored to achieve such goal. This process helps students develop procedural knowledge while ensuring that AI complements their efforts rather than replaces them.

\textbf{Guide students in asking effective questions.} The quality of interaction with an LLM heavily depends on the clarity and specificity of the student's queries. Teachers should provide guidance on how to frame questions that prompt meaningful and educationally valuable responses. For example, rather than asking an LLM to ``solve this equation," students could ask, ``What are the general steps to solve quadratic equations, and why do they work?" This approach encourages critical engagement and deepens conceptual understanding.

\textbf{Promote reflection and critical analysis.} LLMs are not infallible and can produce outputs that are plausible but incorrect or incomplete. Teachers should use these moments as opportunities to teach students how to critically evaluate AI-generated responses. For instance, if an LLM provides a mathematical derivation with an error, students can be encouraged to identify and correct it. This not only builds subject knowledge but also enhances students' comprehension of concepts, rather than focusing on specific technical aspects. Furthermore and indirectly, this improves students’ critical thinking skills and their ability to verify information independently.

\textbf{Enhance group projects and co-creativity.} Teachers can facilitate collaborative projects where students use LLMs to brainstorm ideas, develop strategies, or analyze data. For example, in a science class, a group could use an LLM to design an experiment, asking the model for guidance on setting up variables, controls, and hypotheses. Teachers should emphasize the importance of integrating AI suggestions with students’ ideas, ensuring that AI augments rather than dominates the creative process.

\textbf{Highlight ethical and practical implications.} Finally, it is crucial for teachers to discuss the ethical dimensions of AI use with students, emphasizing responsible practices. For instance, teachers can explain how LLMs generate responses based on training data, highlighting issues like potential biases, privacy concerns, and the importance of intellectual property. Students should be encouraged to use LLMs responsibly and to consider the broader implications of their outputs in real-world contexts.



\section{Conclusion}
\label{sec:con}
As we discussed in this paper, Artificial Intelligence (AI), and particularly Large Language Models (LLMs), have the potential to revolutionize education by enabling personalization, inclusivity, and expanding students horizons. This paper has explored how LLMs, when used effectively, can transform the learning experience, allowing students to progress at their own pace, explore real-world problems, and co-create innovative solutions. By serving as both patient tutors and collaborative partners, LLMs can address diverse educational needs, from filling foundational knowledge gaps to guiding advanced or interdisciplinary projects.

However, to fully realize these benefits, AI must be integrated thoughtfully into the educational process. It is essential that LLMs support students in developing resolving strategies, rather than merely providing answers. This collaborative approach encourages critical thinking, independence, and creativity, ensuring that students remain active participants in their learning journeys. Additionally, educators play a pivotal role in defining how AI tools are used in classrooms. By teaching students how to engage with AI responsibly and critically, teachers can help ensure that these tools are leveraged to enhance, rather than hinder, meaningful learning.
In this paper, we have highlighted practical applications and provided case studies to demonstrate how LLMs can personalize learning, support diverse students, and enhance co-creativity. These examples highlight AI's ability to empower learners across various contexts and levels, preparing them for the challenges of an AI-driven future.

\section*{Acknowledgement}

This work has been funded by the ``Progetti di Terza Missione" of Sapienza University of Rome under grant TM12318B71779C0A, ``YRiA: Young Reserachers in Action". We also would like to thank G.B. Grassi High School of Latina, the headmaster, the teachers, and all the students that participated in this project. Furthermore, we would like to thank Prof. Donato Bini for providing the problem in Section \ref{sec:horizons}, Prof. Eugenio Montefusco and Prof. Caterina Rizzo for the material and the tutoring in Section \ref{sec:creativity}.


\bibliographystyle{IEEEbib}
\bibliography{SPMbib}

\section*{Biographies}

\textbf{Eleonora Grassucci} received the Ph.D. degree in Information and Communication Technologies in 2023 from Sapienza University of Rome, Italy. She is an Assistant Professor at the Department of Information Engineering, Electronics, and Telecommunications of the Sapienza University of Rome. She was awarded the Best Track Manuscript Recognition by the IEEE Circuits and Systems Society in 2022 and the Doctoral Dissertation Award by the INNS in 2023.

\textbf{Gualtiero Grassucci} graduated in mathematics in 1990 from Sapienza University of Rome, Italy. He teaches mathematics and physics at the G.B. Grassi High School of Latina, Italy, where he is the ideator of the project \textit{Research in Action} for the introduction of applied mathematics in high school (selected among the best Italian projects in 2020 - Digital School Award and granted by Sapienza University of Rome). In 2019 he received the Cancellieri Prize for the teaching of mathematics.

\textbf{Danilo Comminiello} received the Ph.D. degree in Information and Communication Engineering in 2012 from Sapienza University of Rome, Italy. He is currently an Associate Professor with the Department of Information Engineering, Electronics, and Telecommunications (DIET) of Sapienza University of Rome, Italy. He is the Vice-Chair of the IEEE Machine Learning for Signal Processing Technical Committee and an elected member of the IEEE Nonlinear Circuits and Systems Technical Committee.

\textbf{Aurelio Uncini} received the Ph.D. degree in Electrical Engineering in 1994 from the University of Bologna, Italy. Currently, he is a Full Professor at the Department of Information Engineering, Electronics, and Telecommunications, where he is teaching Neural Networks, Adaptive Algorithms for Signal Processing and Digital Audio Processing, and where he is the Head of the ``Intelligent Signal Processing and Multimedia'' (ISPAMM) group.

\end{document}